\documentclass[prb,superscriptaddress,aps]{revtex4}
\usepackage{graphicx}
\usepackage{dcolumn}
\usepackage{bm}

\usepackage{color}

\begin{document}


\bigskip\bigskip\bigskip

\title{Propagation of ultrastrong femtosecond laser pulses in
PLASMON-X}

\author{Du\v san Jovanovi\'c}
\email{djovanov@ipb.ac.rs} \affiliation{Institute of Physics,
University of Belgrade, Belgrade, Serbia}

\author{Renato Fedele}
\email{renato.fedele@na.infn.it} \affiliation{Dipartimento di
Scienze Fisiche, Universit\`{a} Federico II and INFN Sezione di
Napoli, Complesso Universitario di M.S. Angelo, via Cintia,
I-80126 Napoli, Italy}

\author{Fatema Tanjia}
\email{tanjia@na.infn.it} \affiliation{Dipartimento di Scienze
Fisiche, Universit\`{a} Federico II and INFN Sezione di Napoli,
Complesso Universitario di M.S. Angelo, via Cintia, I-80126
Napoli, Italy}

\author{Sergio De Nicola}
\email{sergio.denicola@ino.it} \affiliation{Istituto Nazionale di
Ottica - C.N.R., Pozzuoli (NA), Italy} \affiliation{Dipartimento
di Scienze Fisiche, Universit\`{a} Federico II and INFN Sezione di
Napoli, Complesso Universitario di M.S. Angelo, via Cintia,
I-80126 Napoli, Italy}

\centerline {\textsf{O3.205 - 38th EPS Conference on Plasma
Physics, Strasbourg, France, 26 June - 1 July, 2011}}

\bigskip\bigskip

\maketitle






Plasmon-X is a project based in the Frascati INFN laboratories
using the Ti:Sa laser FLAME and electrons' linac SPARC. FLAME
(Frascati Laser for Acceleration and Multidisciplinary
Experiments) has a pulse with maximum energy $E_M= 7$ J, maximum
duration $\tau_M =25$ fs, maximum power $W_M = 250$ TW, wavelength
$\lambda = 800\, {\rm nm}$, and repetition rate $\nu_{rep}= 10$ Hz
(see, f.i., \cite{Prometeus}). The pulse duration $\tau = 25\,{\rm
fs}$ corresponds to the pulse length $L_z = 7.5\,\,\mu{\rm m}$
($\sim 10$ wavelengths within the pulse). An upgrade that includes
the polarization control (S, P, circular) is planned for the near
future \cite{Prometeus}. The plasma density in different Plasmon-X
experiments \cite{NTA,Gizzi} ranges as $n_e = 0.6-1\times
10^{19}\, {\rm cm}^{-3}$, up to $4\times 10^{19}\, {\rm cm}^{-3}$.
Note that $n_e = 10^{19}\, {\rm cm}^{-3}$ corresponds to the
plasma frequency $\nu_{p,e}=\omega_{p,e}/2\pi \simeq 28\,{\rm
THz}$. Thus, the pulse duration is $\tau = 0.7 \,\, T_p$, where
$T_p$ is the plasma period $T_p \equiv 2\pi/\omega_{p,e} =
35\,{\rm fs}$, while the collisionless skin depth, $d_e \equiv
2\pi c/\omega_{p,e} \simeq 10.6\,\,\mu$m, is an order of magnitude
longer than the laser wavelength ($\lambda = 0.8\,\, \mu$m) and
close to the pulse length ($L_z = 7.5\,\,\mu{\rm m}$). The laser
wake field (LWF) accelerator scheme \cite{NTA} envisages that an
electron bunch with the energy of 150 MeV and the transverse
normalized emittance of 1 mm$\cdot$mrad, whose transverse and
longitudinal rms sizes are $5\,\,\mu$m and $2.5\,\,\mu$m,
respectively, is injected in the second bucket of the Langmuir
wave excited by a Ti:Sa pulse delivering 7 J of energy in 30 fs.
The laser pulse with initial waist size of $L_\bot=130\,\,\mu$m
and minimum size of $L_\bot=32.5\,\,\mu$m is guided by a matched
channel profile. Plasma is $9.88$ cm long and its density profile
has a positive and varying slope with starting and ending
densities of $1.5\times 10^{17}\,\,{\rm cm}^{-3}$ and $2.5\times
10^{17} \,\, {\rm cm}^{-3}$.

We derive the nonlinear equations that describe the propagation of
ultrashort laser pulses in a plasma, in the Plasmon-X device. We
consider the interaction of the high frequency electromagnetic and
Langmuir waves, while the acoustic phenomena are disregarded. The
laser electric field is so strong that the electrons achieve
relativistic jitter velocities. The nonlinear effects come mostly
from the interaction between the electromagnetic pump wave with a
Langmuir wave, whose frequency is considerably lower than that of
the electromagnetic (laser) pump. The electrons are regarded as
cold, i.e. the phase velocity of the nonlinear modes involved are
much higher than the electron thermal velocity, see. e.g. Refs.
\cite{sprangle,mahajan,shukla}. The characteristic frequency of
the laser light is sufficiently high and the ions are essentially
immobile, i.e. $n_i = n_0$ and $\vec v_i = 0$. The component of
the Ampere's law perpendicular to the direction of propagation of
the laser beam  and the Poisson's equation have the form
{\small\begin{equation}
\label{amplaw-perp} \frac{\partial^2\vec A_\bot}{\partial t^2}-
c^2\left(\nabla_\bot^2+\frac{\partial^2}{\partial z^2}\right) \vec
A_\bot+\nabla_\bot\frac{\partial\phi}{\partial t} =  \frac{\vec
j_\bot}{\epsilon_0}, \quad
\left(\nabla_\bot^2+\frac{\partial^2}{\partial z^2}\right) \phi =
-\frac{\rho}{\epsilon_0},
\end{equation}}
where $\rho$ and $\vec{j}_\perp$ are the charge density and the
perpendicular component of the current density $\vec{j}$,
respectively, that are calculated as $ \rho = \sum_\alpha q_\alpha
n_\alpha$ and $\vec j = \sum_\alpha q_\alpha n_\alpha \vec
v_\alpha, $ where $q_\alpha$ is the charge of the particle species
$\alpha$, and the hydrodynamic densities $n_\alpha$ and velocities
$\vec v_\alpha$ are calculated from the appropriate hydrodynamic
equation. The electron continuity and momentum equations take the
form
\begin{equation}
\label{cont}\frac{\partial n}{\partial t}+\nabla\cdot\left(n\vec
v\right) = 0, \quad \left(\frac{\partial}{\partial t}+\vec
v\cdot\nabla\right) \vec p = q\left[-\nabla\phi-\frac{\partial\vec
A}{\partial t} + \vec v\times\left(\nabla\times\vec
A\right)\right] ,
\end{equation}
where, for simplicity, the subscript for electrons has been
omitted and $q = -e$. Here $\vec p$ is the electron momentum,
related with the electron velocity $\vec v$ through $ \vec v =
{\vec p}/{m_0\gamma}, $ $m_0$ is the electron rest mass, $\gamma =
(1 +p^2/m_0^2 c^2)^\frac{1}{2}$, and $c$ is the speed of light.
Under the Plasmon-X conditions, the solution is slowly varying in
the reference frame moving with the velocity $\vec e_z\, u$ and we
can use the approximation from \cite{sprangle,mahajan}. Using the
dimensionless quantities $\vec{p} \rightarrow {\vec p}/{m_0 c}$,
$\vec{v}\rightarrow {\vec v}/{c}$, $\phi \rightarrow {q\phi}/{m_0
c^2}$, $\vec{A} \rightarrow {q\vec A}/{m_0 c}$, $n \rightarrow
{n}/{n_0}$, $u \rightarrow {u}/{c}$, $t \rightarrow \omega_{p,e}
t$, ${\vec r\,\,} \rightarrow ({\omega_{p,e}}/{c})( \vec r - \vec
e_z\, u t)$, where $\omega_{p,e}$ is the plasma frequency of
stationary electrons, $\omega_{p,e} = (n_0 q^2/m_0
\epsilon_0)^\frac{1}{2}$. The solution of the hydrodynamic
equations (\ref{cont}) is sought in the almost 1-D, i.e. $
\nabla_\bot \ll {\partial}/{\partial z}, $ quasistatic regime,
i.e. $ {\partial}/{\partial t}\ll u \, {\partial}/{\partial z}$,
and $u$ is adopted close to the speed of light, i.e. $ 1-u\ll 1. $
Then, with the accuracy to the leading order, they are integrated
as
\begin{equation}
\label{bdcontL2}\left(v_z - 1\right)n + 1 = 0, \quad - p_z +\gamma
-1 + \phi = 0, \quad \vec p_\bot + \vec A_\bot = 0,\quad A_z = 0.
\end{equation}
Using the definition for $\gamma$, after some straightforward
algebra, we obtain the dimensionless charge- and current densities
as $
n = [{\left(\phi - 1\right)^2 + {\vec{A_\bot}}^2 + 1}]/{2
\left(\phi - 1\right)^2}$ and $ \vec v_\bot n = {\vec
A_\bot}/({\phi - 1})$, which permits us to rewrite our basic
equations (\ref{amplaw-perp}) as {\small
\begin{equation}\label{waveqenv0} \left[\frac{\partial^2}{\partial
t^2} - 2 u\,\,\frac{\partial^2}{\partial t\,\, \partial z} -
\left(1-u^2\right)\frac{\partial^2}{\partial z^2} - \nabla_\bot^2
+ \frac{1}{1-\phi}\right]\vec A_\bot =
-\left(\frac{\partial}{\partial t} - u\,\,
\frac{\partial}{\partial z}\right)\nabla_\bot\phi,
\end{equation}
\begin{equation}\label{poissons0}
\frac{\partial^2\phi}{\partial z^2} = \frac{\left(\phi -
1\right)^2 - 1 - {\vec{A_\bot}}^2}{2 \left(\phi - 1\right)^2}.
\end{equation}
} The above equations constitute a Zakharov-like description of a
modulated electromagnetic wave, coupled with a Langmuire wave via
the nonlinearities that arise from the relativistic effects. Thus,
besides the standard nonrelativistic three-wave coupling phenomena
(the Raman scattering), in the relativistic case there arises also
a possibility for the four-wave processes, that may lead to the
modulational instability, soliton formation, etc. The actual
nonlinear dynamics of the pulse strongly depends on the physical
conditions in each particular device and can not be generalized.
We apply our Eqs. (\ref{waveqenv0}) and (\ref{poissons0}) to the
Plasmon-X conditions, for {\em moderately focussed} laser beams.
With the present power of the laser in the Plasmon-X device, in
most experimental setups we have ${\vec{A_\bot}}^2 < 1$, i.e we
can take the pulse to have a moderate intensity and expand into
series the nonlinear terms in the above equations. We seek the
solution in the form of a modulated electromagnetic wave, viz. $
\vec A_\bot = \vec A_{\bot_0} \,\, \exp\{{-i[\omega' t - k'(
z+ut)]\}} + c.c., $ where the dimensionless frequency $\omega'$
and the dimensionless wavenumber $k'$ are defined as $ \omega'=
{\omega}/{\omega_{p,e}}$, $k' = {ck}/{\omega_{p,e}} =
{d_e}/{\lambda}$, where $\omega$, $k$, and $\lambda$ are the
frequency, the wavenumber, and the wavelength of the
electromagnetic laser wave, respectively, while $\omega_{p,e}$ and
$d_e$ are the electron plasma frequency and the collisionless skin
depth. They satisfy the linear dispersion relation of
electromagnetic waves, $\omega = \sqrt{c^2 k^2 + \omega_{p,e}^2}$,
whose dimensionless version has the form $ \omega = \sqrt{k^2 + 1}
. $ Here and in the following, for simplicity, we drop the primes.
We adopt $u$ to be equal to the group velocity of the
electromagnetic wave $ u = {d\omega}/{d k} = {k}/{\omega}. $
Substituting these into the wave equation and the Poisson's
equation, dropping the nonresonant zero- and double pump frequency
terms in  Eq. (\ref{waveqenv0}) (the latter is absent for a
circularly polarized laser wave), these are further simplified to
{\small
\begin{equation}\label{waveqenvC}
\left[2i\,\omega\,\frac{\partial}{\partial t} +
\frac{1}{\omega^2}\frac{\partial^2}{\partial z^2} + \nabla_\bot^2
- \phi\right]\, A_{\bot_0} = 0,\quad
\left(\frac{\partial^2}{\partial z^2} + 1\right) \phi =
-\frac{\left|A_{\bot_0}\right|^2}{2}.
\end{equation}
{
We studied the 2-D evolution of a moderately focussed pulse,
typical for the Plasmon-X accelerator scheme, by the numerical
solution of Eqs. (\ref{waveqenvC}) in the 2-D regime
$\nabla_\bot^2 = \partial^2/\partial x^2$. The initial condition
was adopted in the form of an unchirped NLS soliton, modulated in
the perpendicular direction by a Gausssian, $A_{\bot_0}(x,z,0)
=2\,\sqrt{C_1}\,\, \exp{(-x^2/2 L_x^2)}\,\, \exp{(i\,\delta k\,
z)}\, {\rm sech}(\sqrt{C_1}\,z)$, with $L_x = 25$, $C_1 = 0.07$,
and $\delta k = 0.5$, with $\phi\left(x,z,0\right) = 0$. We
followed its evolution until $t= 15$. During this time, the pulse
travels approximately 4.5 cm, which is about half the length of
the interaction chamber. The results are displayed in Fig. 1. The
folding of the pancake-like pulse and the creation of a V shape
was observed after $t\sim 5$ and the  simultaneous longitudinal
stretching due to the negative chirp. The related diminishment
affected mostly the laser envelope, while the electrostatic
potential still featured a sizable amplitude; the depth of its
first minimum was more than 50\% of its largest value, achieved
immediately after the launch of the pulse. The potential minimum
obtained a V-shape in the $x$-$z$ plane and its perpendicular
contraction due to collapse was rather small.
\begin{figure}[htb]
\includegraphics[width=78mm]{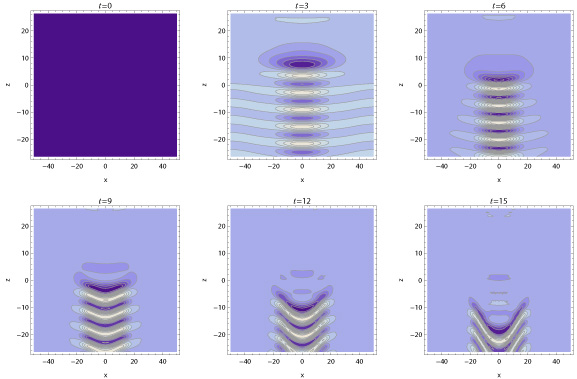}
\includegraphics[width=78mm]{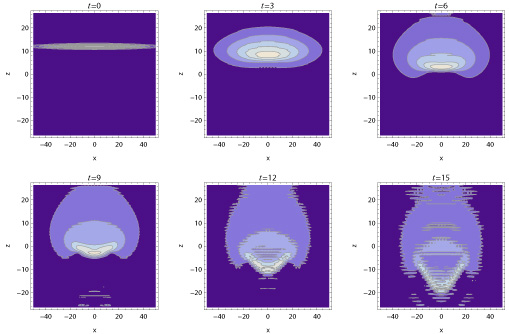}
\caption{\small The electrostatic potential (left) and the
amplitude (right) of a moderately focussed laser pulse, typical
for the Plasmon-X accelerator scheme.} \label{fig9}
\end{figure}

In conclusion, we have shown that the experimental scheme for the
LWF electron acceleration, with the laser power of $W\leq 250$ TW,
the pulse duration $\tau\geq\,\,{\rm2 5 fs}$, and the spot size of
$130\,\,\mu{\rm m}$, can be described by a nonlinear Schr\"
odinger equation with a reactive nonlocal nonlinear term, that
produces an oscillating electrostatic wake to the laser pulse.
Such moderately focussed pulses develop a chirp, that is roughly
proportional to the intensity. The pulse spreads, relatively
slowly, in the direction of propagation and its amplitude
diminishes, and simultaneously a non-selfsimilar transverse
collapse takes place. We do not expect that the latter constitutes
a critical limitation for the Plasmon-X accelerator scheme. For
very large amplitudes, $\sim 30$ times bigger than the potential
in Fig. 1, the collapse is quenched by the saturation of the
nonlinear term in the wave equation. The presence of the chirp
offers a new venue for the stabilization of the collapse even at
smaller amplitudes, known in nonlinear optics and in Bose-Einstein
condensate, and it permits one to introduce some sort of external
control, such as the appropriate profile of the plasma density, to
stabilize the structure.

\bigskip\bigskip

\noindent \textbf{Acknowledgements.} This work was partially
supported by the grant 171006 of the Serbian MSE and partially by
Fondo Affari Internazionali of INFN.


\begin{thebibliography}{99}

\bibitem{Prometeus}
G. Turchetti {\it et al}, \textit{Optical acceleration activity in
Italy:
Plasmonx, Prometheus project},\\
http://wwwapr.kansai.jaea.go.jp/pmrc$\underline{\,\,\,}$en/org/colloquium/download/colloquium16-1.pdf  

\bibitem{NTA}
C. Benedetti {\it et al}, {\it NTA PlasmonX},
http://www.lnf.infn.it/rapatt/2008/06/PLASMONX.pdf 

\bibitem{Gizzi}
L. A. Gizzi  {\it et al}, Eur. Phys. J. Spec. Top., {\bf 175}, 3
(2009).

\bibitem{sprangle}
P. Sprangle, E. Esarey, and A. Ting,
Phys. Rev. Lett., {\bf 64}, 2011 (1990).

\bibitem{mahajan}
V. I.  Berezhiani  and S. M. Mahajan,
Phys. Rev. Lett. {\bf 73}, 1837 (1994).

\bibitem{shukla}
A. Sharma, I. Kourakis, and P. K. Shukla, Phys. Rev. {\bf E 82},
016402 (2010).

\end{thebibliography}

\end{document}